\documentclass{PoS}

\usepackage{graphicx}
\usepackage{amsmath}
\usepackage{amssymb}
\usepackage{color}

\newcommand{\be}{\begin{equation}}
\newcommand{\ee}{\end{equation}}
\newcommand{\bea}{\begin{eqnarray}}
\newcommand{\eea}{\end{eqnarray}}
\newcommand{\bi}{\begin{itemize}}
\newcommand{\ei}{\end{itemize}}
\newcommand{\ben}{\begin{enumerate}}
\newcommand{\een}{\end{enumerate}}
\newcommand{\bt}{\begin{tabbing}}
\newcommand{\et}{\end{tabbing}}

\title{
   $|V_{ud}|$, $|V_{us}|$, $|V_{cd}|$, $|V_{cs}|$
   and charm (semi)leptonic decays:
   WG1 summary from CKM 2016
}

\ShortTitle{
   $|V_{ud}|$, $|V_{us}|$, $|V_{cd}|$, $|V_{cs}|$
   and charm (semi)leptonic decays
}

\author{
   Takashi Kaneko
   \\
   High Energy Accelerator Research Organization (KEK),
   Ibaraki 305-0801, Japan 
   \\
   School of High Energy Accelerator Science,
   SOKENDAI (The Graduate University for Advanced Studies),
   Ibaraki 305-0801, Japan
   \\
   E-mail: \email{takashi.kaneko@kek.jp}
}

\author{
   Xiao-Rui Lyu
   \\
   University of Chinese Academy of Sciences, 
   Beijing 100049, China
   \\
   E-mail: \email{xiaorui@ucas.ac.cn}
}

\author{
   Arantza Oyanguren
   \\
   Instituto de Fisica Corpuscular, 
   Centro Mixto Universidad de Valencia - CSIC, 
   Valencia, Spain
   \\ 
   E-mail: \email{Arantza.Oyanguren@ific.uv.es}
}

\abstract{
   We summarize the current status of the determination of
   the CKM matrix elements $|V_{ud}|$ and $|V_{us}|$,
   which is at the precision frontier of CKM phenomenology.
   We also review 
   recent progress on 
   the study of charm (semi)leptonic decays,
   and the determination of $|V_{cd}|$ and $|V_{cs}|$.
}

\FullConference{
   9th International Workshop on the CKM Unitarity Triangle\\
   28  November - 3 December 2016\\
   Tata Institute for Fundamental Research (TIFR), Mumbai, India
}

\begin{document}

\section{Introduction}

The precise determination of the Cabibbo-Kobayashi-Maskawa (CKM) matrix 
elements is one of the most important subjects in the search for 
new physics.
New bounds on violations of CKM unitarity, for instance,
translate into constraints on models beyond the Standard Model (SM),
and may eventually turn up evidence of new physics.
Due to the CKM hierarchy,
the elements $|V_{ud}|$, $|V_{us}|$, $|V_{cd}|$ and $|V_{cs}|$ 
have (sub-)dominant contributions to the unitarity condition 
in the first and second rows.

At present,
the first row condition $|V_{ud}|^2+|V_{us}|^2+|V_{ub}|^2=1$ 
provides the most stringent test of CKM unitarity.
The relevant elements, $|V_{ud}|$ and $|V_{us}|$, 
have been precisely determined from the super-allowed nuclear $\beta$ decays
and $K\!\to\!\pi \ell\nu$ decays, respectively.
An advantage in controlling hadronic uncertainties
lies in the fact that 
these decays proceed through the weak vector current.
The conservation of the vector current (CVC) and 
the non-renormalization theorem imply that 
the relevant matrix elements at zero momentum transfer are known 
in the isospin (SU(3)) limit.
Corrections to the symmetric limit are quadratic 
in a symmetry-breaking parameter $m_{d(s)}\!-\!m_u$~\cite{CVC:BS,CVC:AG}.

It is of course important to explore various different decays. 
Such decays provide not only independent determinations of 
the CKM elements but also complementary probes of new physics.
For instance, the weak axial current contributes to 
the neutron $\beta$ decays, kaon and pion leptonic decays,
which are therefore sensitive to pseudoscalar- and axial-vector-type
new physics interactions.
Inclusive hadronic $\tau$ decays may be sensitive to new physics
that couples primarily to the third generation.

There are a rich variety of charm decay modes,
which are important in the search for new physics. 
The $D_{(s)}\!\to\!\ell\nu$ and $D\!\to\!\pi(K)\ell\nu$ decays 
provide a precise determination of $|V_{cd(s)}|$. 
Rare and forbidden decay modes
may serve as sensitive probes of new physics.
Large data samples of the charm decays are being accumulated 
at charm and $B$ factories leading to 
a recent remarkable improvement in the experimental accuracy. 
Control of the hadronic uncertainties is more difficult
than that in the kaon decays
due to the large flavor symmetry breaking.
Thanks to the continuous development of powerful computers and 
simulation algorithms, 
the accuracy of lattice QCD determinations of the hadronic matrix elements 
has also been improved in recent years.

In this summary, 
we present an overview of recent progress 
reported in the Working Group 1 sessions of the CKM 2016 workshop.


\section{$|V_{ud}$, $|V_{us}|$ and unitarity in the first row}

\subsection{$|V_{ud}|$ from nuclear and neutron decays}



The super-allowed nuclear $\beta$ decays,
namely $0^+\!\to\!0^+$ transitions of isospin-one ($I\!=\!1$) nuclei, 
have provided the most precise determination of $|V_{ud}|$.
A key quantity is the product of 
the phase space factor $f$
and the partial half-life $t$.
By eliminating transition-dependent corrections
$\delta_R^\prime$, $\delta_{NS}$ and $\delta_C$,
we can define a corrected value
\bea
   {\mathcal F}t
   & = &
   ft \left( 1 + \delta_R^\prime)(1 + \delta_{NS} -\delta_C \right)
   =
   \frac{K}{2 G_V^2 \left(1 + \Delta_R \right)},
\eea
which is expected to be transition independent.
Here $\delta_R^\prime$ and $\delta_{NS}$ represent the radiative corrections,
whereas $\delta_C$ is the isospin-symmetry breaking correction.
Note that $\delta_{NS}$ and $\delta_C$ depend on 
the details of nuclear structure.
In the right-hand side,
$\Delta_R$ represents the transition-independent radiative correction.
The vector coupling $G_V$ is related to 
$|V_{ud}|$ and the Fermi coupling $G_F$ as $G_V\!=\!|V_{ud}|G_F$.
Since the constants $K$ and $G_F$ have negligibly small uncertainties,
we can obtain $|V_{ud}|$ 
from experimental determination of $ft$ and
theoretical calculation of $\delta_C$, $\delta_{NS}$, $\delta_R^\prime$ and $\Delta_R$.

\begin{figure}[t]
\begin{center}
   \includegraphics[angle=0,width=1.0\linewidth,clip]{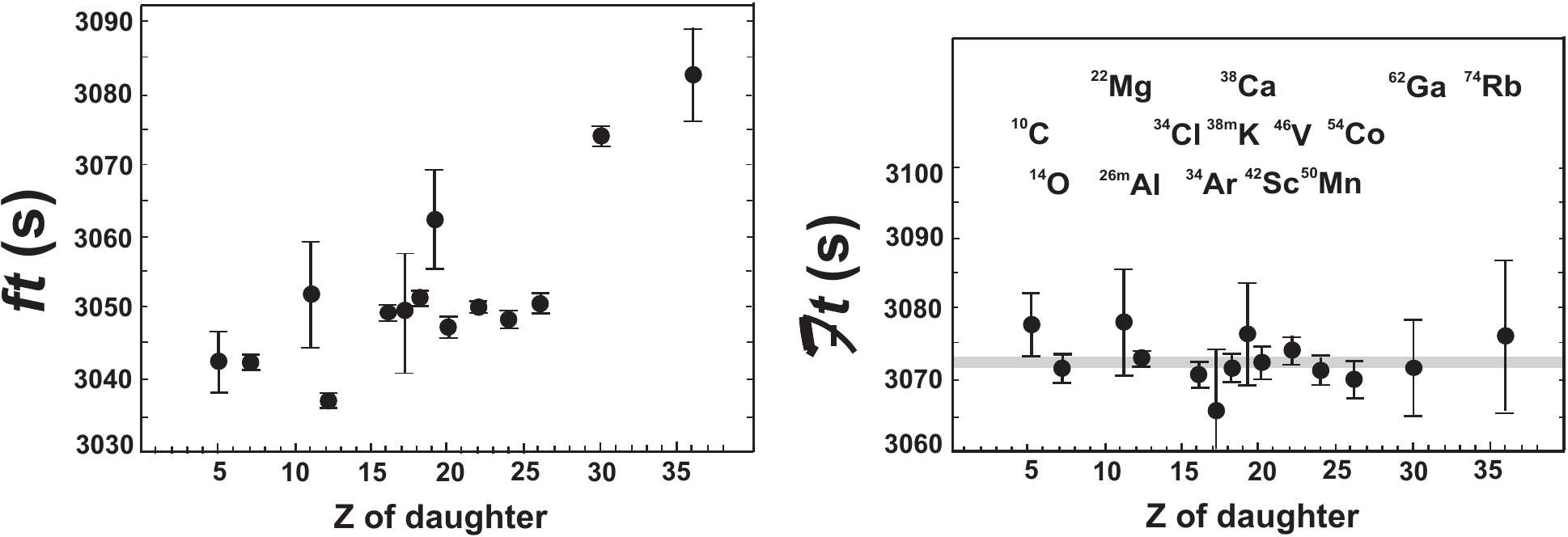}

   \vspace{0mm}
   \caption{
     Results for the uncorrected $ft$ values (left panel) and 
     corrected ${\mathcal F}t$ values (right panel)
     for the 14 best-known super-allowed nuclear $\beta$ decays.
     (Figure from Ref.~\cite{Vud:Hardy:CKM16}.)
   }
   \label{fig:Vud:Ft}
\end{center}
\vspace{-5mm}
\end{figure}

As reviewed by J.C.~Hardy~\cite{Vud:Hardy:CKM16},
new experimental results became available after the last workshop CKM 2014:
the total transition energy and branching ratio
for the decay branch from $^{14}O$,
and the half-life of $^{10}C$~\cite{Vud:Q:14O,Vud:BR:14O,Vud:t:16O}, 
which are inputs to determine $ft$.
The left panel of Fig.~\ref{fig:Vud:Ft} shows the updated average of $ft$ 
for the 14 best-known decays~\cite{Vud:Hardy:CKM16}.
The precision is $\leq\!0.05$\,\% for the nine $ft$ values,
and $<\!0.3$\,\% for the other five cases.


The horizontal axis of the panel represents
the atomic number $Z$ of the daughter nucleus. 
The $Z$ dependence of $ft$ is small (note the vertical axis scale of the figure)
but significant at the high precision of $ft$.
Among the transition-dependent corrections,
the isospin correction $\delta_C$ becomes
importantly large as $Z$ increases.
Several methods have been proposed to calculate $\delta_C$~\cite{Vud:isospin}.
Only the shell-model calculation using the so-called 
Saxon-Woods radial wave-function leads to the impressive agreement
of the corrected ${\mathcal F}t$ values
as shown in the right panel of Fig.~\ref{fig:Vud:Ft}.
This confirms the CVC expectation of a universal value of $G_V$
at the level of $\pm 0.01$\,\%.


The accuracy of ${\mathcal F}t$ is then further improved
by averaging over $Z$.
Theoretical corrections are small (partly) due to CVC.
These decays therefore yields very precise estimate:
\bea
   |V_{ud}|
   & = & 
   0.97420(21)
   \hspace{3mm}
   (\mbox{super-allowed nuclear decays}).
   \label{eqn:Vud:Vud}
\eea
This 0.02\,\% uncertainty is dominated by that from 
the calculation of the transition independent correction $\Delta_R$.
Therefore, 
only little reduction of this uncertainty is possible
without improved calculation of $\Delta_R$.

The neutron $\beta$ decays receive the radiative corrections
$\delta_R^\prime$ and $\Delta_R$,
but they are free of $\delta_{NS}$ and $\delta_C$,
which depend on the nucleus structure.
These decays may therefore be better to determine $|V_{ud}|$ in the long term.
The master formula is
\bea
   |V_{ud}|^2
   & = &     
   \frac{4908.7(1.9)[s]}{\tau_n \left(1+3\lambda^2\right)},
\eea
where $\tau_n$ is the neutron lifetime.
In contrast to the super-allowed nuclear decays,
the neutron decays proceed also through the weak axial current.
These decays therefore provide an independent determination of $|V_{ud}|$
with different sensitivity to new physics.
However, the vector and axial-vector contributions have to be disentangled 
through a difficult decay correlation measurement 
to fix $\lambda\!=\!G_A/G_V$, where $G_A$ is the axial coupling.
A target accuracy is 0.02\,--\,0.03\,\% both for $\tau_n$ and $\lambda$
to be competitive to the determination from the super-allowed nuclear decays.
The current status and future prospect are summarized by D.~Po\v{c}ani\'c
in Ref.~\cite{Vud:Pocanic:CKM16}.


\begin{figure}[t]
\begin{center}
   \includegraphics[angle=0,width=0.50\linewidth,clip]{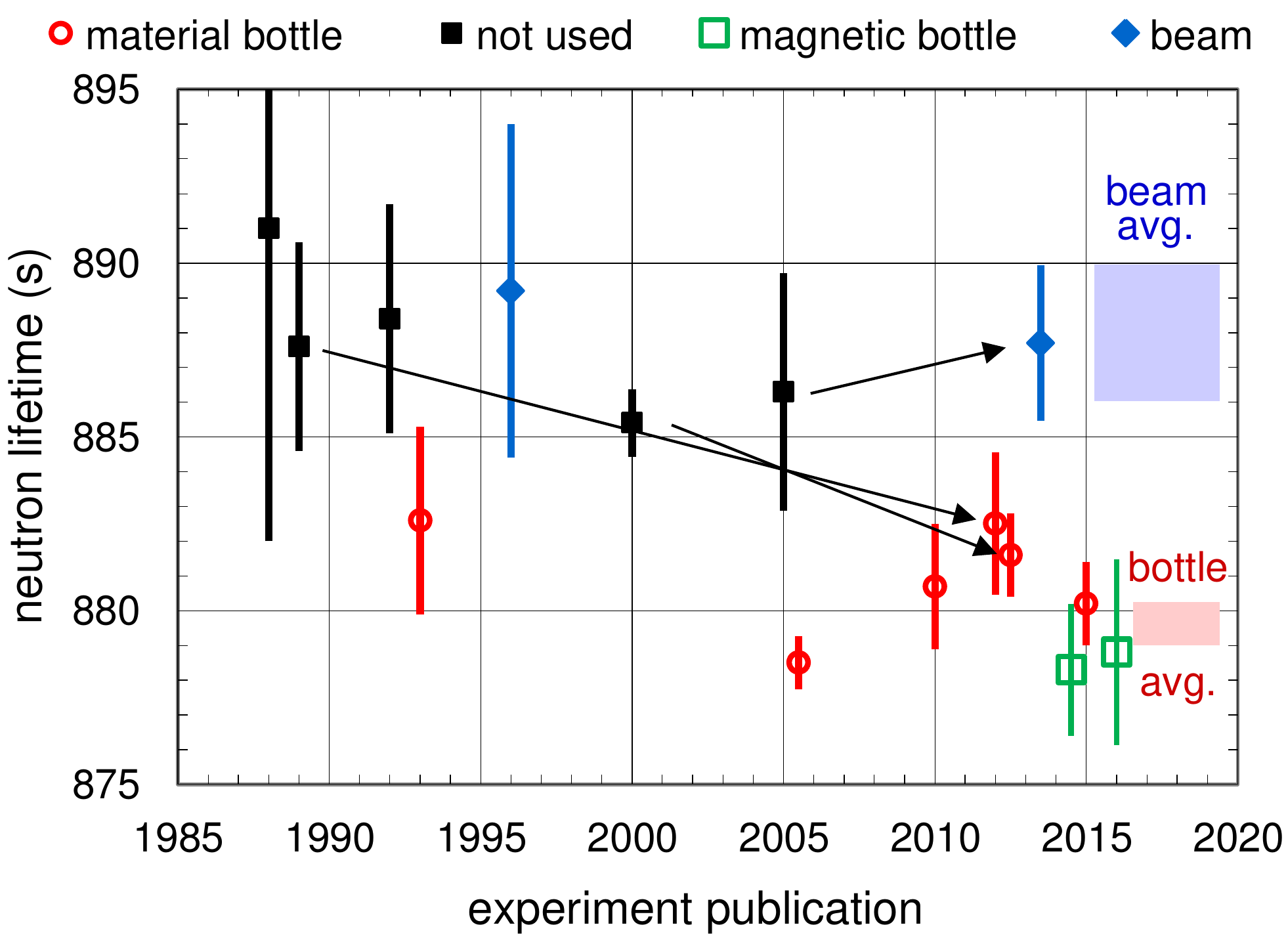}
   \hspace{5mm}
   \includegraphics[angle=0,width=0.40\linewidth,clip]{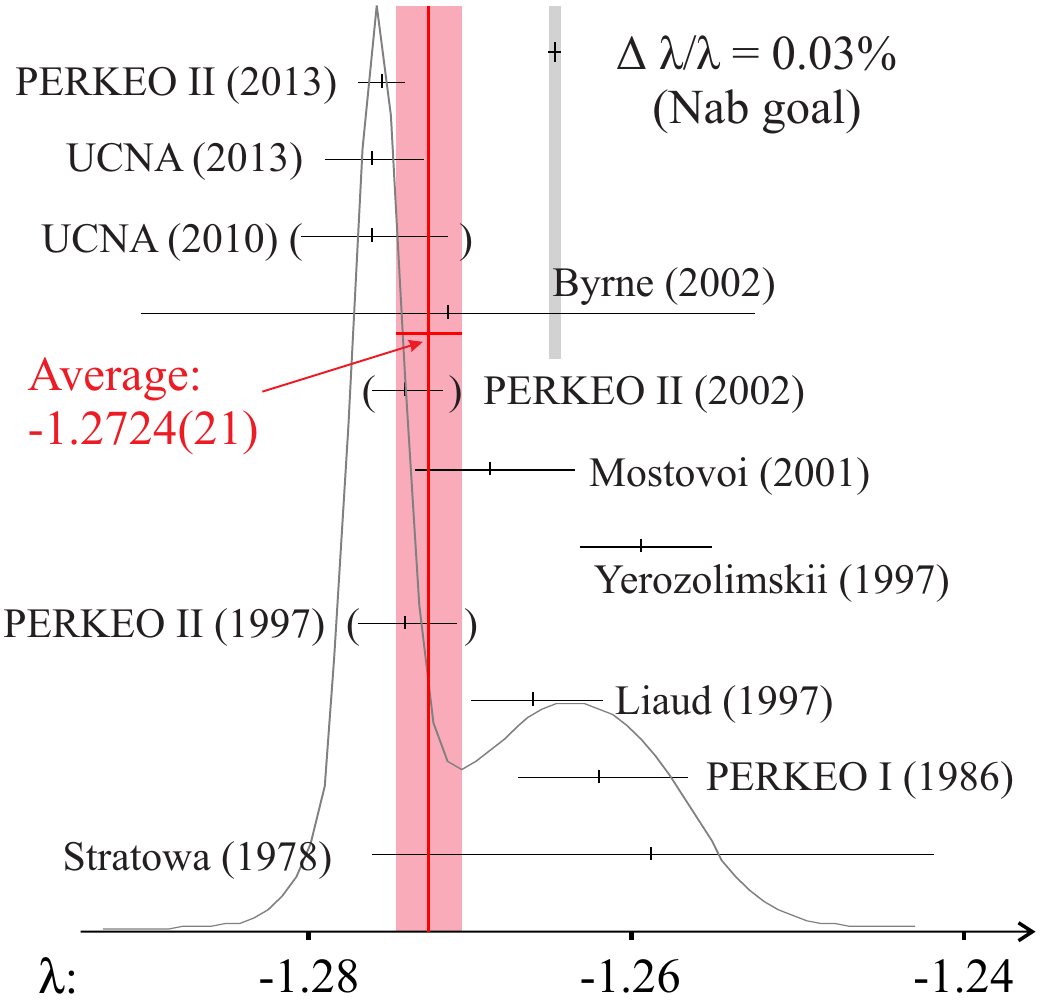}

   \vspace{0mm}
   \caption{
      Left panel:
      neutron lifetime measurements using the beam method (diamonds)
      and bottle method (circles and squares).
      Two shaded regions show the averages of $\tau_n$ from two methods.
      Right panel:
      measurements of $\lambda$. The red band shows its global average,
      whereas the planned precision of the upcoming Nab experiment~\cite{Vud:Nab}
      is shown in the gray band.
      (Figure from Ref.~\cite{Vud:Pocanic:CKM16}.)
   }
   \label{fig:Vud:neutron}
\end{center}
\vspace{-5mm}
\end{figure}


The neutron lifetime $\tau_n$ has been measured by two methods~\cite{Vud:neutron:lifetime}.
In the {\it beam} method~\cite{Vud:CN}, 
the number of neutron decays is counted
as a cold neutron beam passes through a fiducial volume.
On the other hand, 
the {\it bottle} method~\cite{Vud:UCN} stores ultra-cold neutrons
in a material or magneto-gravitational bottle,
and counts the survivors after some storage time.
As shown in the left panel of Fig.~\ref{fig:Vud:neutron}, however,
averages of $\tau_n$ from these two methods
appear to be systematically different from each other:
888.0(2.0)\,$s$ and 879.6(0.6)\,$s$ from the beam and bottle methods,
respectively.
Underestimated systematic uncertainties or unknown effects 
are the most likely cause of this $4\,\sigma$ discrepancy,
which is to be understood by forthcoming experiments~\cite{Vud:Pocanic:CKM16}.


The ratio $\lambda$ can be independently extracted from
three correlation coefficients in the neutron decay rate~\cite{Vud:neutron}:
the $\beta$ asymmetry $A$,
$\nu_e$ asymmetry $B$,
and $e$-$\nu_e$ correlation coefficient $a$.
It is known that $A$ and $a$ have comparable sensitivities to $\lambda$,
while $B$ is much less sensitive.
So far, $A$ has been measured more precisely than $a$,
and yields the most accurate results for $\lambda$.
However, 
the world data of $\lambda$
in the right panel of Fig.~\ref{fig:Vud:neutron}
are not perfectly consistent among themselves
and lead to the global average $\lambda\!=\!-1.2724(21)$
with a poor confidence level. 

These inputs yields $|V_{ud}|\!=\!0.9758(16)$
with significantly larger uncertainty compared to Eq.(\ref{eqn:Vud:Vud}).
As listed in Ref.~\cite{Vud:Pocanic:CKM16}, however, 
there are many on-going and planned measurements both for $\tau_n$ and $\lambda$
to resolve the discrepancies among different methods and
to improve the accuracy of $|V_{ud}|$.


%

\subsection{$|V_{us}|$ from kaon and $\tau$ decays}



The $K\!\to\!\pi \ell\nu$ semileptonic decays, namely the $K_{l3}$ decays, 
provide a precise determination of $|V_{us}|$.
The decay rate is given by
\bea
   \Gamma_{K_{\ell3}}
   & = & 
   \frac{G_F^2 M_K^5}{192\pi^3} C_K^2 
   I_{K\ell} 
   S_{\rm EW} \left( 1 + \delta_{\rm EM}^{K\ell} + \delta_{\rm SU(2)}^{K\pi} \right)^2
   |V_{us} f_+^{K^0\pi^-}(0)|^2,
   \label{eqn:Vus:Kl3:Gamma}
\eea
where $C_K$ is a Clebsch-Gordan coefficient and 
$I_{K\ell}$ is the phase-space integral. 
The short-distance electroweak correction is given by $S_{\rm EW}\!=\!1.0232(3)$.
The long-distance electromagnetic (EM) and isospin corrections
are denoted 
by $\delta_{\rm EM}^{K\ell}$ and $\delta_{\rm SU(2)}^{K\pi}$, respectively,
and have been estimated based on chiral perturbation theory (ChPT).
The relevant form factors are defined from the hadronic matrix element 
\bea
   \langle \pi(p^\prime) | V_\mu | K(p) \rangle
   & = & 
   \left\{
      p + p^\prime - \frac{M_K^2-M_\pi^2}{q^2} q
   \right\}_\mu f_+^{K\pi}(q^2)
  +\frac{M_K^2-M_\pi^2}{q^2} q_\mu f_0^{K\pi}(q^2),
   \label{Vus:FF}
\eea
where $q^2\!=\!(p^\prime-p)^2$ is the momentum transfer.
Only the vector component of the weak current contributes
to these decays, and SU(3) breaking effects to $f_+^{K\pi}(0)$
are second order in $m_s-m_u$~\cite{CVC:AG}.


The kaon leptonic decays ($K_{\ell2}$), on the other hand,
proceed through the weak axial current. 
The relevant hadronic input, the decay constant $f_K$,
breaks SU(3) invariance already at first order.
This and other corrections can partially cancel 
in the decay rate ratio to the pion decays ($\pi_{\ell2}$)
\bea
   \frac{\Gamma_{K_{\ell2}}}{\Gamma_{\pi_{\ell2}}}
   & = & 
   \frac{M_K\left(1-m_{\ell}^2/M_K^2\right)^2}{M_\pi\left(1-m_{\ell}^2/M_\pi^2\right)^2}
   \left( 1 + \delta_{\rm EM} \right) 
   \frac{|V_{us}|^2}{|V_{ud}|^2}
   \frac{f_K^2}{f_\pi^2},
   \label{eqn:Vus:Pl2:Gamma}
\eea
where $\delta_{\rm EM}$ is the long-distance EM correction.
These $P_{\ell2}$ ($P\!=\!K, \pi$) decays therefore provide
an independent determination of $|V_{us}|/|V_{ud}|$~\cite{Vus:Pl2}.


The experimental inputs, $\Gamma_{\{K_{\ell3},K_{\ell2}\}}$ and $I_{K\ell}$,
were precisely measured by
kaon experiments between 2003 and 2010~\cite{Vus:Kl3:Flavia}.
After the last workshop,
there are no significant new experimental inputs 
with full error budget.
Only the ChPT estimate of $\delta_{\rm SU(2)}^{K\pi}$
has been slightly changed
with updated inputs (quark mass ratios) from lattice QCD~\cite{FLAG3}.
The values $|V_{us}f_+^{K\pi}(0)|\!=\!0.21654(41)$ and
$|V_{us}| f_K / |V_{ud}| f_\pi \!=\! 0.27599(37)$
remain essentially unchanged~\cite{Vus:Moulson:CKM16}.
As reviewed by M.~Moulson in detail, 
there are good prospects for a wealth of new experiments.
For instance, the 0.19\,\% uncertainty of $|V_{us}f_+^{K\pi}(0)|$
may be reduced to $\approx 0.12$\,\%
within the next five years~\cite{Vus:Moulson:CKM16}.


\begin{figure}[t]
  \hspace*{-5mm}
   \includegraphics[angle=0,width=0.51\linewidth,clip]{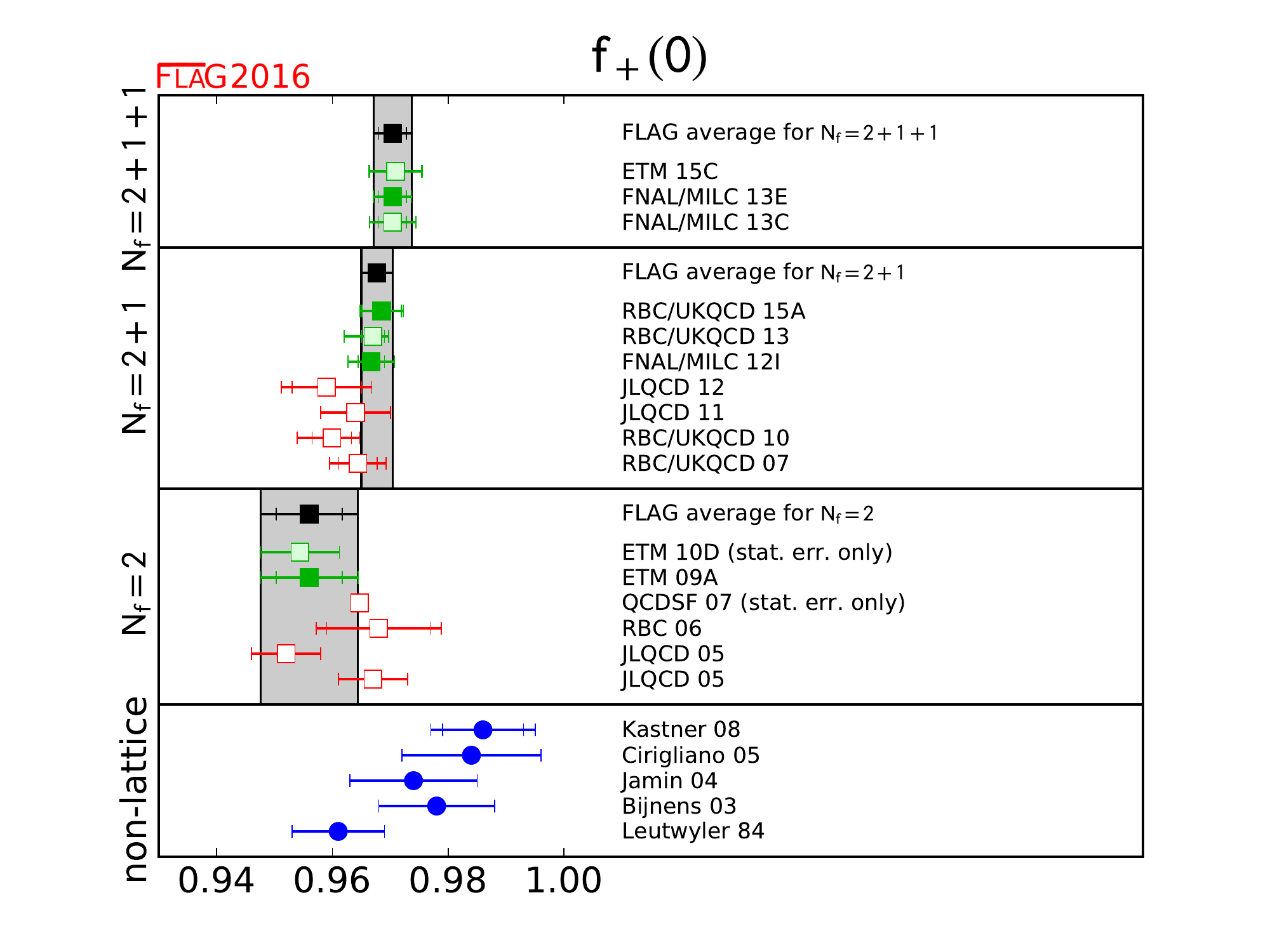}
   \hspace{-7mm}
   \includegraphics[angle=0,width=0.57\linewidth,clip]{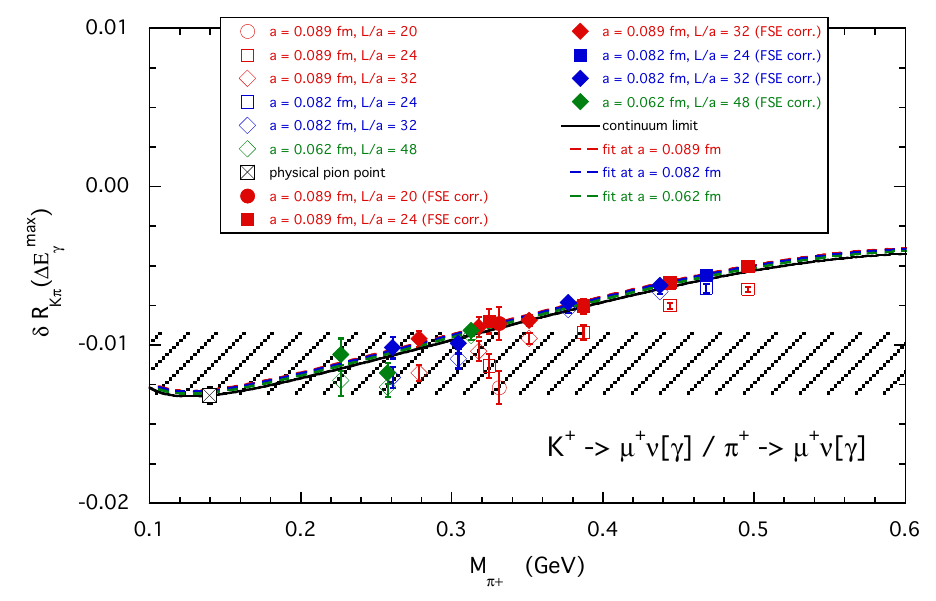}

   \vspace{-2mm}
   \caption{
     Left panel:
     lattice determination of $f_+^{K\pi}(0)$ in $N_f\!=\!2$, 2+1 and 2+1+1 QCD
     (squares) and phenomenological estimate (circles).
     The black square shows the average for each $N_f$.
     (Figure from Ref.~\cite{FLAG3}.)
     Right panel:
     EM and isospin correction to the decay rate ratio~(\protect\ref{eqn:Vus:Pl2:Gamma}).
     Lattice simulation results are plotted by symbols,
     and lines are their fit curves
     at finite lattice spacings and in the continuum limit.
     The shaded band represents the ChPT estimate~\cite{Vus:Pl2:EM+SU2:ChPT:CN,Vus:Pl2:EM:PDG}.
     (Figure from Ref.~\cite{Vus:Pl2:EM+SU2:LQCD}.)
   }
   \label{fig:Vus:Kl3}
\vspace{-5mm}
\end{figure}

As emphasized by S.~Simula~\cite{Vus:Pl2:EM+SU2:LQCD:CKM16},
the accuracy of the lattice QCD determination of the hadronic inputs
has been steadily improved by large-scale simulations
near (or even at) the physical quark masses on fine and large lattices.
The left panel of Fig.~\ref{fig:Vus:Kl3} presents 
a compilation of such realistic simulations.
The current world averages for $N_f\!=\!2+1+1$ QCD, 
$f_K/f_\pi\!=\!1.1933(29)$~\cite{FLAG3}
and $f_+^{K\pi}(0)\!=\!0.9706(27)$~\cite{FLAG3:web},
have now the total uncertainty of 0.3\,\%,
and yield
\bea
   |V_{us}|
   & = &
   0.2231(9),
   \hspace{3mm}
   \frac{|V_{us}|}{|V_{ud}|}
   = 0.2313(6)
   \hspace{3mm}
   (\mbox{$K_{\ell3}$ and $P_{\ell2}$ decays}).
   \label{eqn:Vus:Vus}
\eea
%


At the impressive accuracy of the hadronic inputs,
the uncertainty of the EM and isospin corrections,
which is typically 0.1\,--\,0.4\,\%~\cite{Vus:Kl3:Flavia},
is no longer negligible.
%
%
It is difficult to extend the ChPT calculation to higher
orders where many additional unknown low energy constants appear.
Recently,
a new strategy was proposed to calculate the EM correction
on the lattice for hadronic processes, 
where infrared divergences are present~\cite{Vus:EM+SU2:LQCD}.
This has been succesfully applied
to the $P_{\ell2}$ decays~\cite{Vus:Pl2:EM+SU2:LQCD}:
their preliminary estimate of $\delta_{\rm EM}$ plus
the isospin correction to $f_K/f_\pi$ is
$\delta R_{K\pi}\!=\!-0.0137(13)$,
which is in good agreement with the ChPT estimate
$-0.0112(21)$~\cite{Vus:Pl2:EM+SU2:ChPT:CN,Vus:Pl2:EM:PDG}
as shown in the right panel of Fig.~\ref{fig:Vus:Kl3}.
%


\FIGURE{
   \label{fig:Vus:tau}
   \includegraphics[angle=-90,width=0.45\linewidth,clip]{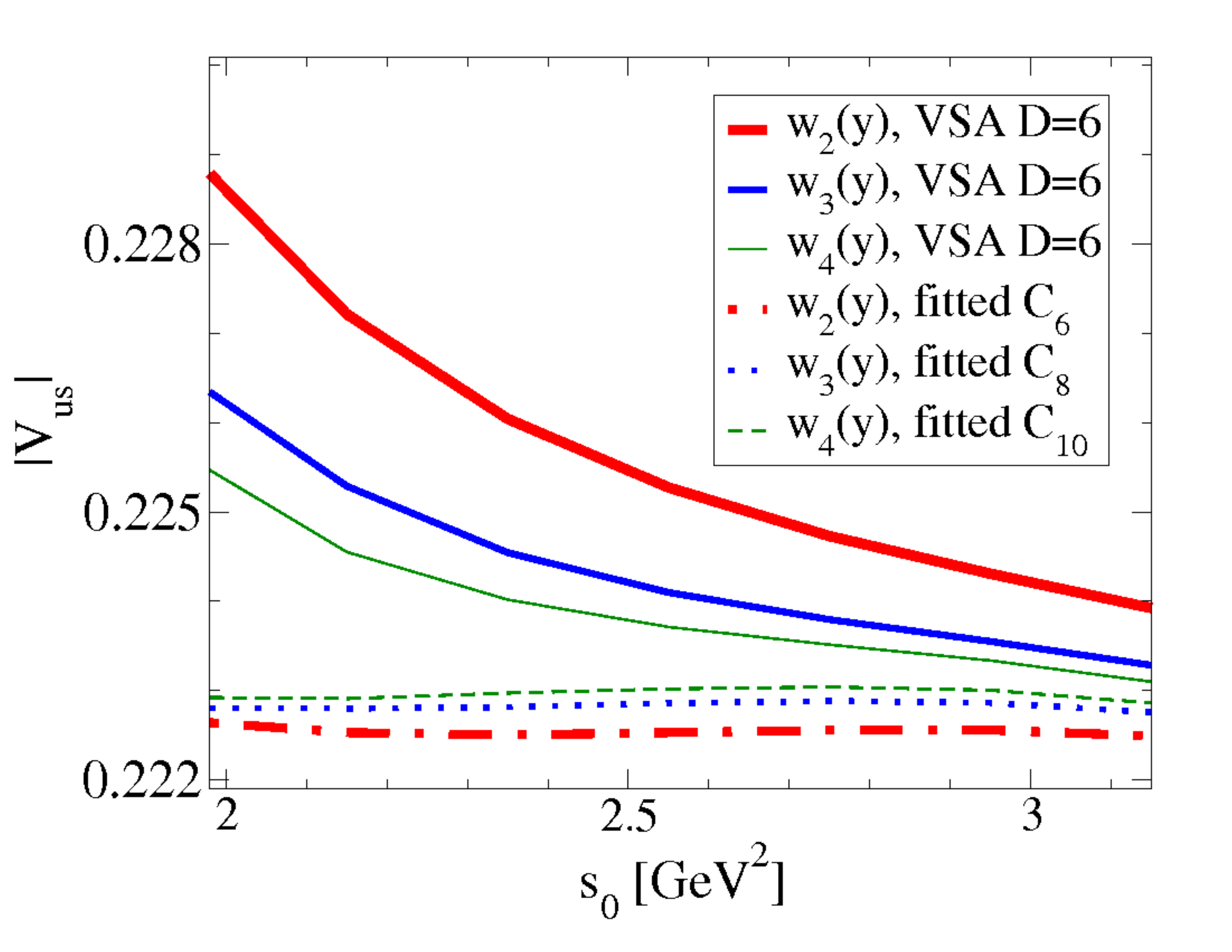}
                   
   \vspace{0mm}
   \caption{
     Estimate of $|V_{us}|$ from hadronic $\tau$ decays
     as a function of $s_0$. 
     Three different weight functions $w_{\{2,3,4\}}$ are 
     tested in the conventional (solid lines) 
     and new implementations (dot-dashed, dotted and dashed lines).
     (Figure from Ref.~\cite{Vus:tau:MHLWZ}.)
   }
}
The inclusive hadronic $\tau$ decays offer 
an alternative determination of $|V_{us}|$~\cite{Vus:tau}.
Previous estimate at the time of the CKM 2014 workshop is 
$|V_{us}|\!=\!0.2176(21)$~\cite{HFAG14},
and about 3~$\sigma$ below the value quoted in Eq.~(\ref{eqn:Vus:Vus}).
This determination employs the finite energy sum rule
to estimate the partial inclusive decay rate
$\Gamma(\tau\!\to\!X_{\{s,d\}}\nu_\tau)$. A key relation is 
\bea
   && 
   \int_0^{s_0} ds\, w(s) \Delta \rho_\tau(s)
   \nonumber 
   \\ 
   &&
   \hspace{10mm}
 =-\frac{1}{2\pi i} \oint_{|s|=s_0} ds\, w(s) \Delta \Pi_\tau(s),
   \hspace{10mm}
   \label{eqn:Vus:FESR}
\eea
where $s$ denotes the momentum transfer to the hadronic state $X_{\{s,d\}}$.
The spectral function $\Delta \rho_\tau$ is experimentally accessible
from the differential distribution of the $\tau$ decays,
and hence contains information on $|V_{us}|$.
The hadronic vacuum polarization function $\Delta \Pi_\tau$ is evaluated 
by the operator product expansion (OPE). 
(Here, ``$\Delta$'' indicates the difference between $\tau\!\to\!X_s\nu_\tau$
and $X_d\nu_\tau$ decays.)
In the conventional set up~\cite{Vus:tau},
the weight function is set to the kinematical factor
$w(s)\!=\!(1-y)^2(1+2y)$ ($y\!=\!s/m_\tau^2$), $s_0\!=\!m_\tau^2$,
and the vaccum saturation approximation (VSA) is assumed 
to evaluate the higher order corrections in the OPE.
However, Eq.~(\ref{eqn:Vus:FESR}) should be valid for any $s_0$
and analytic $w(s)$.

K.~Maltman presented a detailed study of 
theoretical uncertainties~\cite{Vus:Maltman:CKM16}.
As shown in Fig.~\ref{fig:Vus:tau},
$|V_{us}|$ estimated in the conventional setup
significantly depends on the choice of $w(s)$ and $s_0$.
This dependence is largely reduced in a new implementation,
in which both $|V_{us}|$ and non-perturbative parameters at higher orders 
in the OPE are fixed from experimental inputs~\cite{Vus:tau:MHLWZ}.
Together with preliminary BaBar estimate of 
BR($\tau\!\to\!K^-\pi^0\nu_\tau$)~\cite{Vs:tau:BaBar},
the new implementation yields $|V_{us}|\!=\!0.2229(22)_{\rm ex}(4)_{\rm th}$~\cite{Vus:Maltman:CKM16},
which is in good agreement with Eq.~(\ref{eqn:Vus:Vus}).

Future improvements of the experimental inputs are important 
to be competitive with the determination from the kaon decays.
The current experimental status is summarized by S.~Banerjee~\cite{Vus:Banerjee:CKM16}.
Another strategy is proposed in Refs.~\cite{Vus:Maltman:CKM16,Vus:tau:MHLIOZ,Vus:tau:Ohki}
to use designed weights and lattice QCD data of 
the vacuum polarization function.


\subsection{CKM unitarity in the first row}

\FIGURE{
   \label{fig:CKM_1stR}
   \includegraphics[angle=0,width=0.4\linewidth,clip]{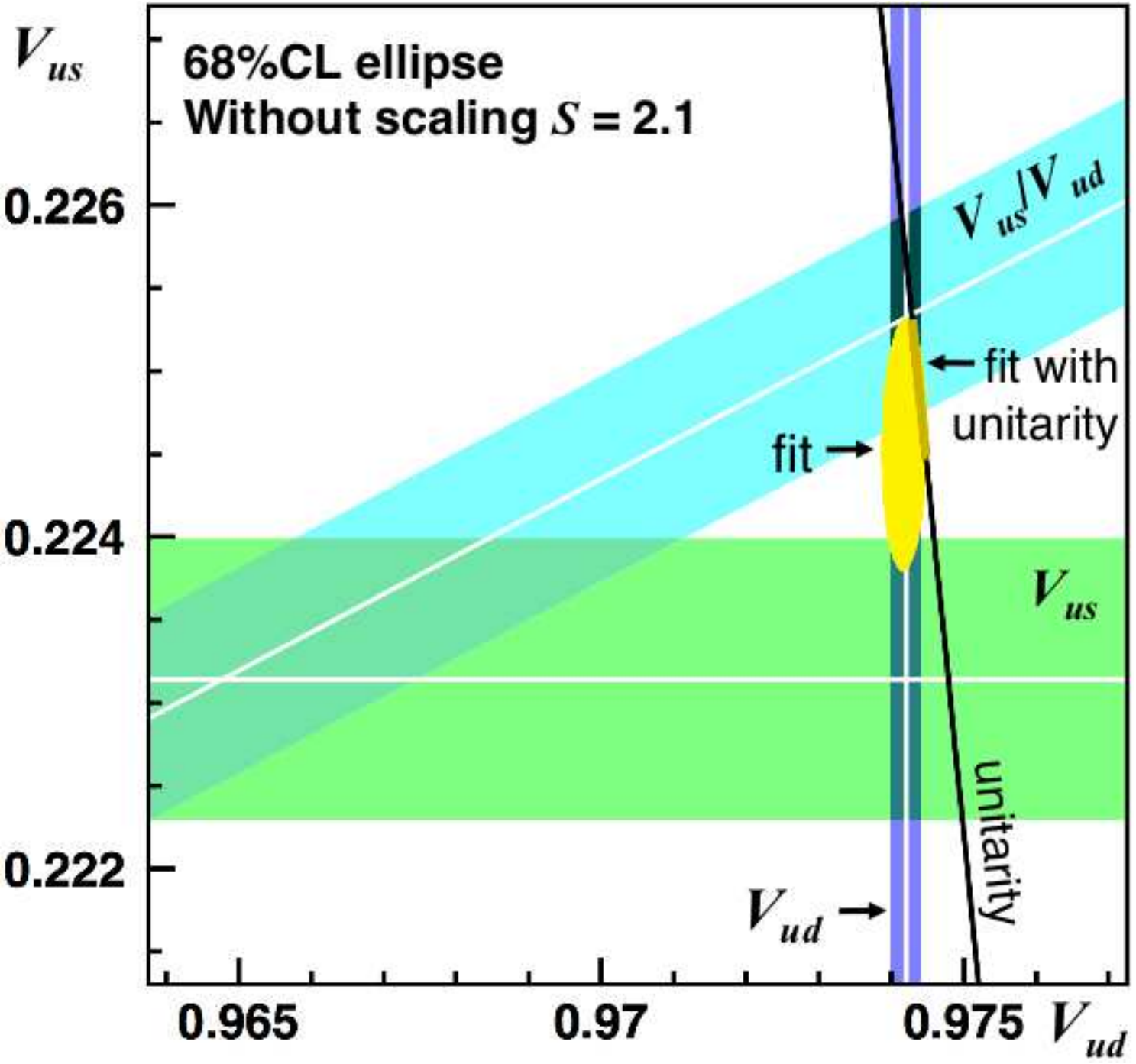}
                   
   \vspace{0mm}
   \caption{
     Test of CKM unitarity in the first row.
     The yellow region is obtained from a single fit to
     $|V_{ud}|$, $|V_{us}|$, $|V_{us}|/|V_{ud}|$ from
     the nuclear, $K_{\ell3}$ and $P_{\ell2}$ decays
     (inputs from $N_f\!=\!2+1+1$ lattice QCD are used for the latter two).
     This region is consistent with the black solid line,
     which satisfies CKM unitarity.
     (Figure from Ref.~\cite{Vus:Moulson:CKM16}.)
   }
}
Reference~\cite{Vus:Moulson:CKM16} combines
$|V_{ud}|$ from the super-allowed nuclear decays,
$|V_{us}|$ and $|V_{us}|/|V_{ud}|$ from the $K_{\ell3}$ and $P_{\ell2}$ decays
in a single fit.
With the hadronic inputs from $N_f\!=\!2+1+1$ lattice QCD,
this fit yields 
$|V_{ud}|\!=\!0.97418(21)$ and $|V_{us}|\!=\!0.2246(5)$,
which are consistent with Eqs.~(\ref{eqn:Vud:Vud}) and (\ref{eqn:Vus:Vus}).
A measure of the CKM unitarity violation is estimated as 
\bea
   \Delta_{\rm CKM}
   & = &
   |V_{ud}|^2+|V_{us}|^2+|V_{ub}|^2-1
   \nonumber
   \\
   & = & 
   -0.0005(5),
   \label{eqn:CKM_1stR:dCKM}
\eea
where $|V_{ub}|\!\approx\!4\!\times\!10^{-3}$ has negligibly small contribution.
%
%
This does not change significantly ($\Delta_{\rm CKM}\!=\!-0.0006(5)$)
when we employ the lattice input for $N_f\!=\!2+1$.
The current determination of $|V_{ud}|$ and $|V_{us}|$, therefore, confirms
the unitarity in the first row at 0.1\,\% accuracy. 
%
%
Model independent analyses based on effective field theory suggest that
this precision test is sensitive to new physics with typical scale of
$\lesssim 10$~TeV~\cite{NP:1st_row:CJGA,NP:1st_row:GAC}.


\section{$|V_{cd}|$, $|V_{cs}|$ and charm (semi)leptonic decays}

\subsection{Leptonic decays}

Large data samples of charm decays have been collected by CLEO-c and BESIII 
at the $\psi(3770)$ and by the $B$ factories, 
Belle and BaBar, at the $\Upsilon(4S)$. 
Branching fractions of leptonic decays of $D_q$ mesons are expressed as
\bea
   \mbox{BR}(D_q^{+}\to \ell^+\nu)
   & = &
   \frac{G_{F}^{2}}{8\pi} \tau_{D_q} f^{2}_{D_q}
   |V_{cq}|^2 M_{D_q} m_{\ell}^2 \bigg( 1 - \frac{m^{2}_{\ell}}{M^{2}_{D_{q}}}\bigg)^2,
\label{eq:charm_lep}
\eea
where $\tau_{D_q}$ and $f_{D_q}$ represent the lifetime and decay constant
of $D_q$ meson, respectively.

\begin{figure}[t]
\begin{center}
   \includegraphics[angle=0,width=0.45\linewidth,clip]{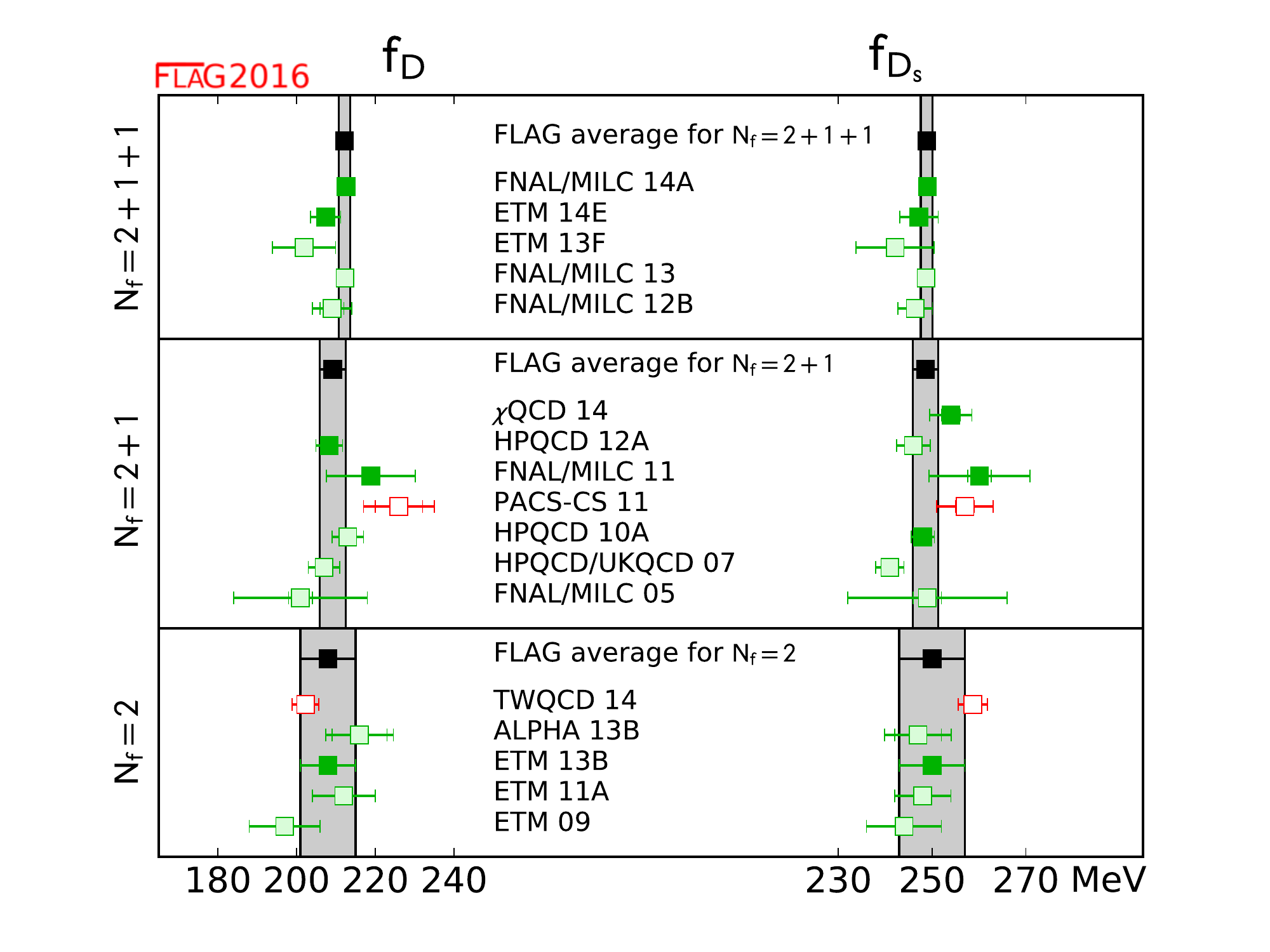}
   \includegraphics[angle=0,width=0.45\linewidth,clip]{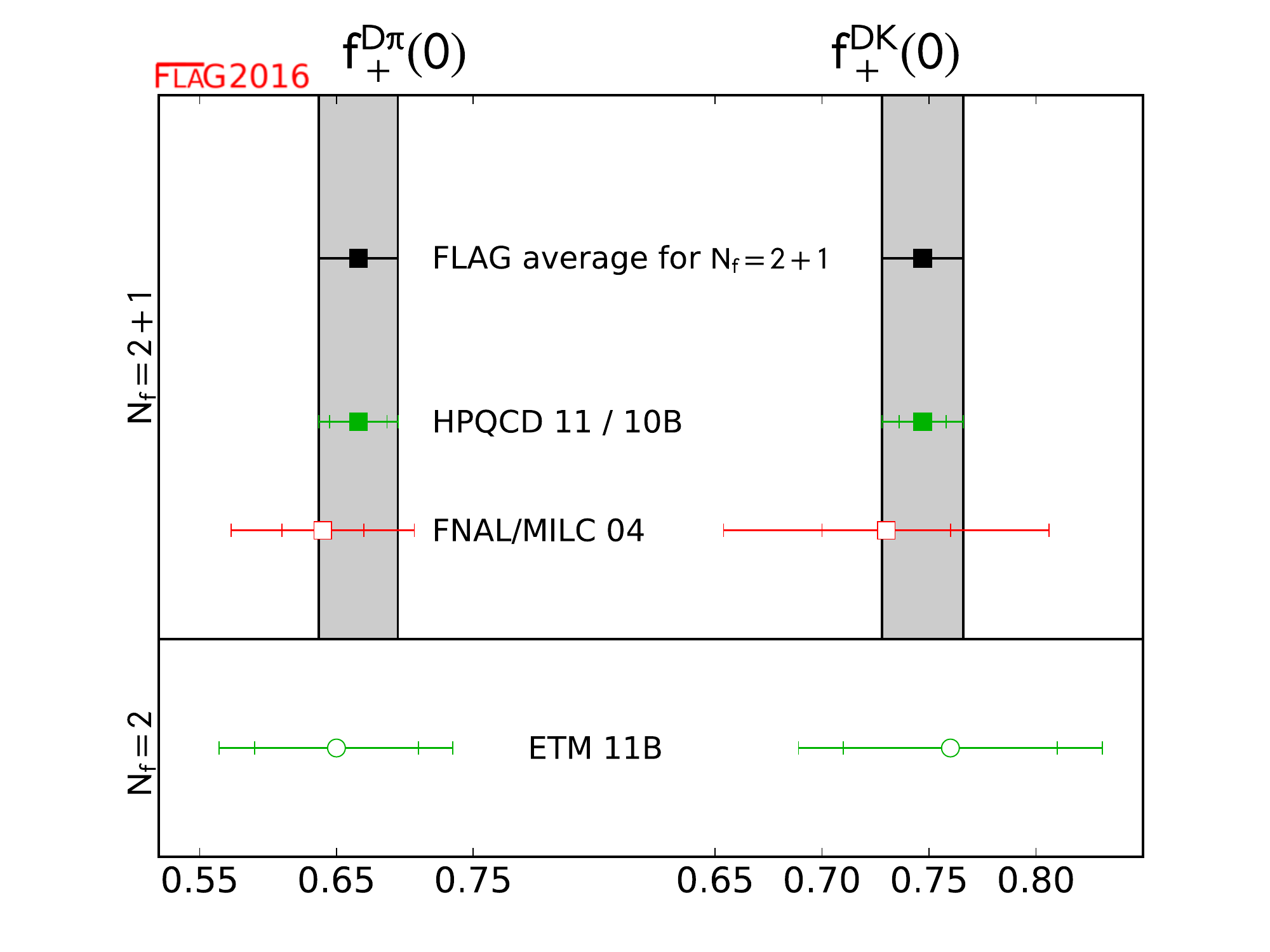}
                   
   \vspace{0mm}
   \caption{Lattice QCD calculations of decay constants $f_{D_{(s)}}$ (left panel)
and $D\!\to\!K(\pi)$ semileptonic form factors $f_+^{DK(\pi)}(0)$ at $q^2 = 0$ (right panel). (Figure from Ref.~\cite{FLAG3}.)
   \label{fig:charm_lattice}}
\end{center}
\vspace{-5mm}
\end{figure}

As discussed by J.T.~Tsang~\cite{D:Tsang:CKM16},
the lattice QCD determination of $f_{D_q}$ has been largely improved 
by realistic simulations near the physical quark masses
on fine lattices. In particular, 
the latter enables us to use relativistic charm quark actions 
with good control of discretization errors and renormalization of $f_{D_q}$.
As shown in the left panel of Fig.~\ref{fig:charm_lattice}, 
there have been independent calculations 
with different lattice actions.
The accuracy of the world average is 0.7 (0.5)\,\% for $f_D(f_{D_s})$.
As in the case of the kaon decays,
the isospin and EM corrections start to be relevant 
at this level of accuracy.
Reference~\cite{D:leptonic:SU(2):FNAL/MILC}
presented a lattice QCD estimate of the isospin correction
$f_{D^+}-f_{D^0}\!=\!0.47\left(^{+25}_{-6}\right)$~MeV.
We also note that 
the method of Ref.~\cite{Vus:EM+SU2:LQCD} can be also applied to 
the $D_q$ meson decays.

H.~Ma reviewed recent experimental progress~\cite{D:Ma:CKM16}.
A measurement of the BR($D^+\to \mu^+ \nu$) has been performed by BESIII using 2.93~fb$^{-1}$ $\psi(3770)$ data set, from the recoil of tagged $D^-$ mesons \cite{Ablikim:2013uvu}. The result has 5\% accuracy, which is dominated by the statistical uncertainty from about 400 signal events. The accuracy of this measurement will be improved significantly if 10 fb$^{-1}$ additional data is taken by BESIII. Preliminary result of the first measurement of BR($D^+\to \tau^+ \nu$) has  also been presented by BESIII, which gives a ratio BR($D^+\to \tau^+ \nu$)/BR($D^+\to \mu^+ \nu$) = $3.21 \pm 0.64$, compatible with the SM prediction.
Results for BR($D_s\to \ell^+ \nu$) have been obtained in the $\mu$ and $\tau$ channels by BaBar and Belle using the full data sample
\cite{delAmoSanchez:2010jg,Zupanc:2013byn}, and by BESIII using 482~pb$^{-1}$ data at center-of-mass energy of 4.009~GeV~\cite{Ablikim:2016duz}. The world average values are BR($D_s\to \mu^+ \nu$) = $(5.54 \pm 0.23)\!\times\!10^{-3}$ and BR($D_s\to \tau^+ \nu$) =  $(5.51 \pm 0.24)\!\times\!10^{-2}$ \cite{HFAG16}. Improved measurements are expected from Belle II and BESIII using more data in the near future. 

\subsection{Semileptonic decays}

The $D\!\to\!P\ell\nu$ semileptonic decays, where $P$ represents
the final state pseudoscalar meson,
proceed through the weak vector current.
The relevant matrix element is parameterized as in Eq.~(\ref{Vus:FF}).
The contribution from the scalar form factor $f_0^{DP}$ is suppressed
by the lepton mass squared $m_\ell^2$, and can be neglected for light leptons.
The differential decay rate is given by 
\bea
\frac{d\Gamma(D\to P \ell \nu)}{dq^2}=\frac{G_{F}^{2}}{24\pi^3}p_{P}^3|f_+^{DP}(q^2)|^2|V_{cq}|^2,
\eea
where $p_P$ is the momentum of $P$ in the $D$ rest frame. 
The $q^2$ dependence of $f_+^{DP}$ has been often
parameterized as a sum of effective poles, 
including the lowest-lying $c\bar q$ resonance with appropriate quantum numbers.
It is popular in recent analyses to use the so-called $z$-parameter expansion~\cite{D:z-param}, which is a model independent parameterization based on
the analyticity of $f_+^{DP}$.
Experiments determine the parameters describing the form factor shape (effective pole masses, coefficients of $z$-parameter expansion) and the normalization of the form factor at $q^2=0$ times the CKM matrix element, namely $f_+^{DP}(0) |V_{cq}|$.

\begin{figure}[t]
\begin{center}
   \includegraphics[angle=0,width=0.45\linewidth,clip]{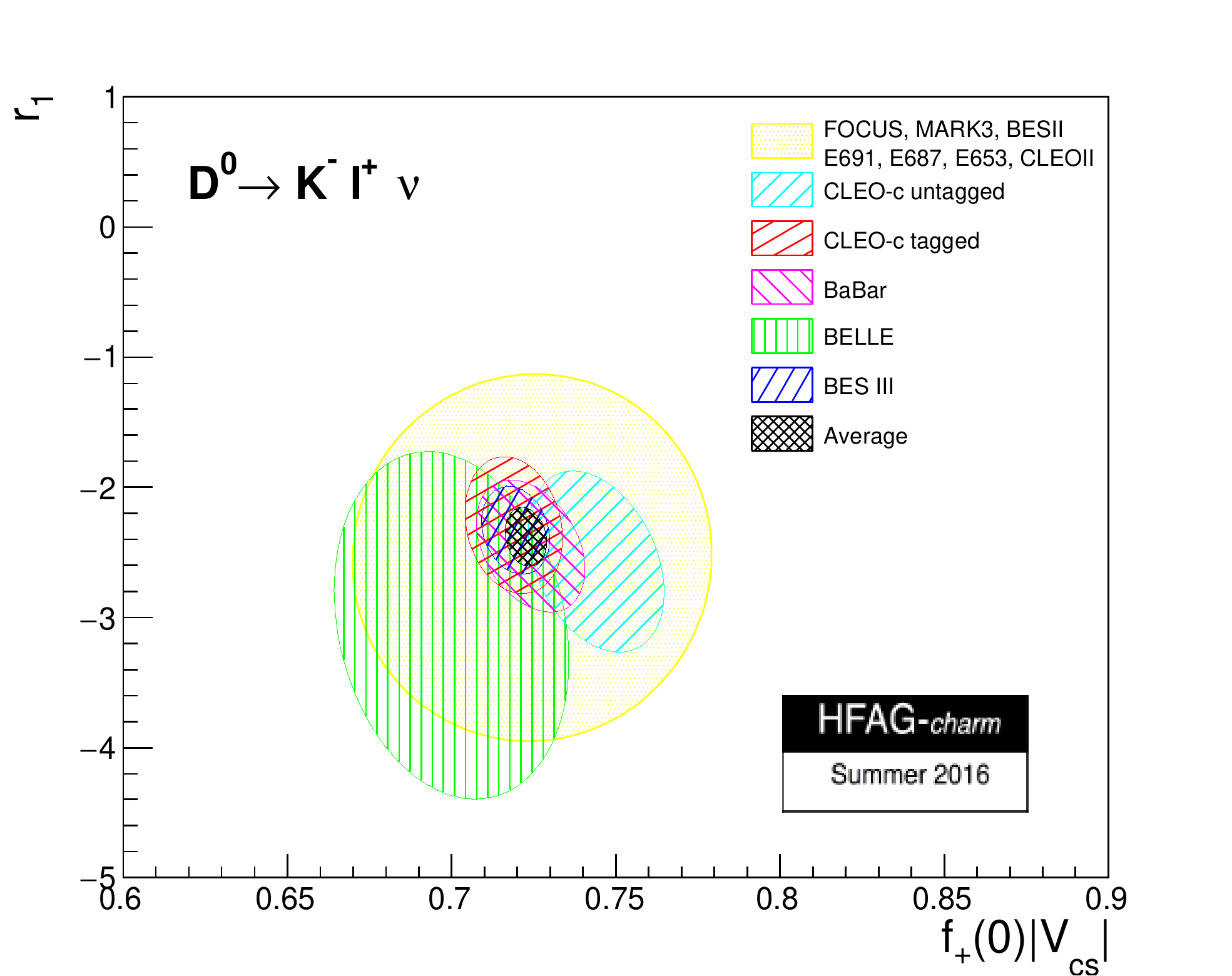}
   \includegraphics[angle=0,width=0.45\linewidth,clip]{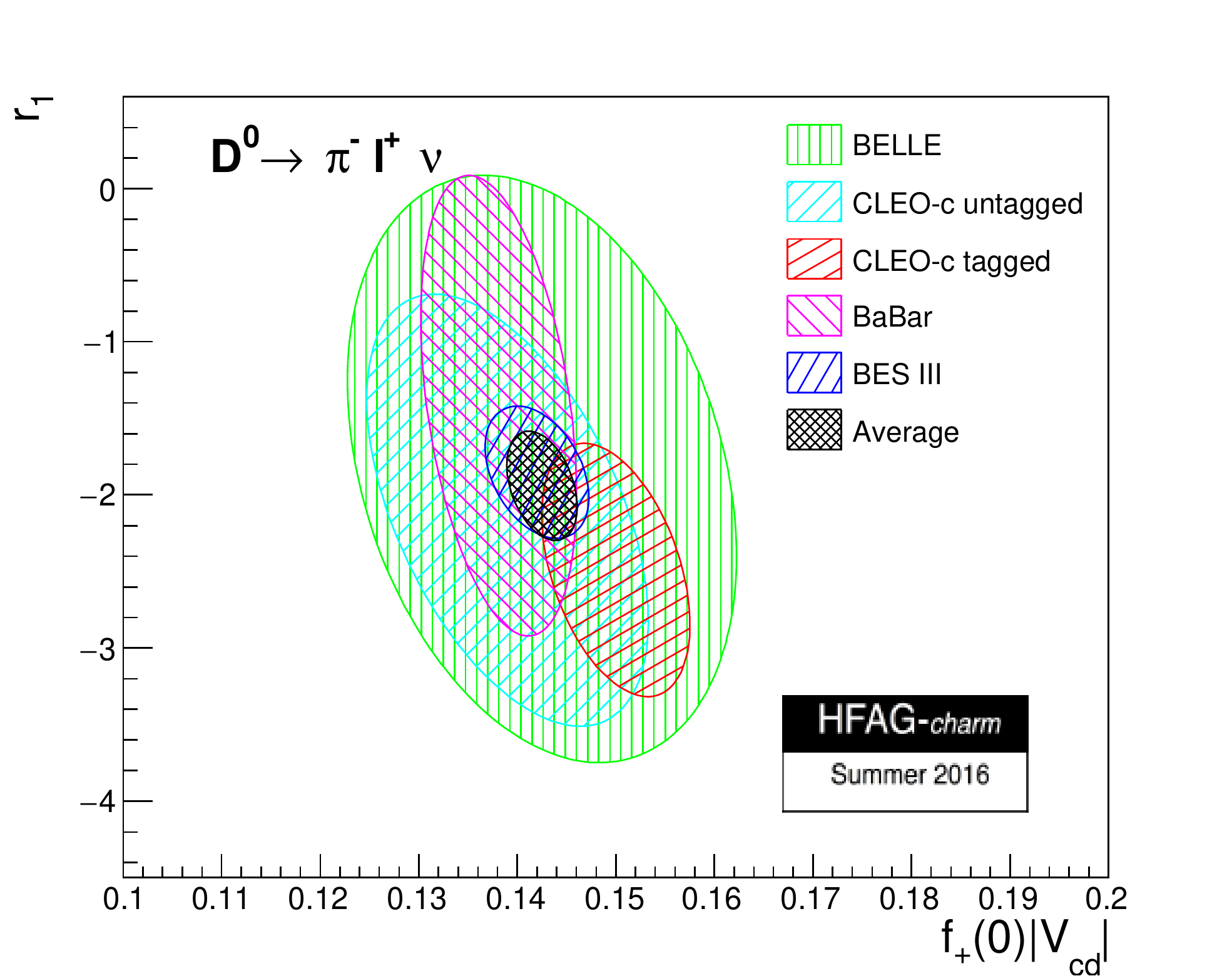}

   \vspace{1mm}
   \caption{Experimental results for $f_+^{DP}(0) |V_{cq}|$ (horizontal axis)
     and a ratio of coefficients in $z$-parameter expansion (vertical axis).
     The left and right panels are for the 
     $D \to K \ell \nu $ and $D \to \pi \ell \nu $ channels, respectively.
     Results obtained by different experiments and the HFAG averages
     are plotted. (Figure from Ref.~\cite{HFAG16}.)
     }
   \label{fig:charm_sl}
\end{center}
\vspace{-3mm}
\end{figure}

Recent experimental progress is reviewed by
A.~Soffer~\cite{D:Soffer:CKM16} and Y.~Zheng~\cite{D:Zheng:CKM16}.
Figure~\ref{fig:charm_sl} compares the results from different experiments~\cite{HFAG16}. The BaBar experiment has recently analysed the $D^0 \to \pi^- e^+ \nu $ channel~\cite{Lees:2014ihu}. The most precise values for the $D^0 \to K^- \ell^+ \nu $ and $D^0 \to \pi^- \ell^+ \nu $ channels are obtained by the BESIII experiment using 2.93~fb$^{-1}$ $\psi(3770)$ data set~\cite{Ablikim:2015ixa}. 
Combined results of measured form factors by several experiments have reached an accuracy of 
$0.5\%$ and $1\%$ for the Cabibbo-allowed and suppressed modes, respectively \cite{HFAG16}.  
Results for the $D^+$ channel, with $K_S$, $K_L$ and $\pi^0$ in the final state, have also been reported by BESIII, although they are less precise than the ones obtained for $D^0$ channel. Results support isospin conservation. 
Other interesting modes with a vector in the final state, $D \to \{K^*, \omega, \phi \} \ell \nu$, and $D_s \to \{\phi,\eta,\eta'\} \ell \nu$ are also being measured~\cite{D:Zheng:CKM16}. 
Furthermore, BESIII performed the first absolute measurement of BR($\Lambda_c^+ \to \Lambda \ell^+ \nu$) based on 0.567~fb$^{-1}$ data  at 4.6~GeV (near the $\Lambda_c^+\bar{\Lambda}_c^-$ mass threshold), which has precision of 12~\%. We expect 10 times more $\Lambda_c^+$ data in the coming years at BESIII and the precision reduced to being less than 4~\%.

As discussed by A.~Davis~\cite{D:Davis:CKM16},
the LHCb experiment could also contribute in the near future in measuring charm semileptonic branching fractions and CKM matrix elements. With the 3~fb$^{-1}$ collected in $pp$ collisions during Run-I, about $5\times 10^{6}$ of signal events are expected for the Cabibbo-allowed mode. Neutrino reconstruction in this environment is challenging, but information on the $D$ flight direction and track momenta can be used to constraint the neutrino transverse momentum in a similar manner as in $b$-hadron semileptonic decays \cite{Aaij:2015bfa}.

In contrast to the calculation for the kaon decays
(Fig.~\ref{fig:Vus:Kl3}),
there have been only few lattice estimates of the 
the form factor normalization $f_+^{DP}(0)$
as shown in the right panel of Fig.~\ref{fig:charm_lattice}.
The accuracy is roughly 3\,\%~\cite{D:FF:DK:HPQCD,D:FF:Dpi:HPQCD},
which is much larger than the experimental accuracy for $f_+^{DP}(0)|V_{cq}|$.
However, the lattice determination can be straightforwardly improved
by more realistic simulations with relativistic charm quark actions.
Indeed, as reviewed by E.~G\'amiz~\cite{D:Gamiz:CKM16},
there are independent on-going lattice calculations~\cite{D:FF:ETM,D:FF:JLQCD,D:FF:FNAL/MILC}.
In addition, 
lattice QCD data are available over the whole $q^2$ region~\cite{D:FF:q2:HPQCD}, and become more precise towards the zero recoil limit. 
Therefore, the accuracy of $|V_{cq}|$ from the semileptonic decays
is expected to be significantly improved 
by a global fit of all experimental and lattice data
using a model independent parametrization,
such as the $z$-parameter expansion.
We also note that a first lattice calculation of
the $\Lambda_c\!\to\!\Lambda$ semileptonic form factors became available~\cite{D:FF:Lambda_c}. These baryonic decays may offer an independent determination
of $|V_{cs}|$ in the future.

\subsection{$|V_{cd}|$, $|V_{cs}|$ and search for new physics}

With the above-mentoned progress,
the CKM matrix elements
\bea
  |V_{cs}|
  & = & 0.997(17),
  \hspace{3mm}
  |V_{cd}|=0.216(5)
\eea
are extracted from the charm leptonic and semileptonic decays~\cite{HFAG16}.
This confirms the unitarity in the second row with an accuracy of 7\,\%
(2\,$\sigma$)
\bea
  |V_{cd}|^2 + |V_{cs}|^2 + |V_{cb}|^2 
  & = &
  1.042(34),
\eea
where $|V_{cb}|\!\approx\!4\!\times\!10^{-2}$ has only small effect.

As discussed by S.~Fajfer~\cite{D:Fajfer:CKM16},
present accuracy of the theoretical and experimental inputs
offers the possibility to search for new physics in $c \to s$ transitions, by looking for small deviations 
from the SM predictions in several observables such as branching ratios,  forward-backward asymmetry in $D\to K \ell\nu $ decays
and transversal muon polarization \cite{Barranco:2014bva,Fajfer:2015ixa}.
In addition, 
tests of lepton flavor universality in charm decays are also proposed~\cite{Fajfer:2015ixa,deBoer:2015boa} that use the BR($D^+\to \pi^+ \mu^+ \mu^-$)/BR($D^+\to \pi^+ e^+ e^- $) ratio in different $q^2$ bins. 
The SM prediction for this observable has a per mille accuracy in the range of [1.25-1.73]GeV$^2$. Predicted limits for lepton flavor violation
$c \to u \ell\ell^\prime$ are also available~\cite{deBoer:2015boa}.


\section{Conclusions}

As reported in the  WG1 sessions,
there have been important theoretical and experimental progress 
in the determination of $|V_{ud}|$,  $|V_{us}|$,  $|V_{cd}|$ and $|V_{cs}|$.
CKM unitarity is now confirmed with 0.1\,\% and 7\,\% precisions
in the first and second rows, respectively.

The accuracy of $|V_{ud}|$ and $|V_{us}|$ is rather stable 
in recent years. 
However, reliability has been steadily improved by, for instance,
a thorough test of the isospin corrections to the super-allowed nuclear decays,
and by resolving the long-standing puzzle on $|V_{us}|$ 
from the inclusive hadronic $\tau$ decays.
A long-standing challenge towards a more stringent unitarity test
is improving the calculation of the transition-independent radiative correction 
to the nuclear decays.
Recent remarkable progress in precision lattice calculations of
kaon matrix elements may justify renewed experimental efforts
on the kaon (semi)leptonic decays~\cite{Plenary:Ceccucci:CKM16}.

Experiments and lattice QCD are in a healthy competition
towards high precision study of charm decays 
leading to recent rapid improvemenet in the determination of 
$|V_{cd}|$ and $|V_{cs}|$.
There are good prospects for future experimental progress
by BESIII, Belle II and possibly LHCb.
In the near future, 
we expect significant improvement 
in the lattice determination of the semileptonic charm decay form factors.
The accuracy of other hadronic inputs for $K$ and $D$ decays
are now below 1\,\%. At this level, isospin and EM corrections
have to be taken into account in a controlled way.
The lattice QCD determination of these correction
is under active development.
\vspace{3mm}

The work of TK is supported in part by the Grant-in-Aid of the MEXT
(No.~26400259)
and by MEXT as ``Priority Issue on Post-K computer''
(Elucidation of the Fundamental Laws and Evolution of the Universe) and JICFuS.


\end{document}